\documentclass[journal]{IEEEtran}
\usepackage{generic}
\usepackage{url}
\usepackage{amsmath,amssymb,amsfonts}
\usepackage{graphicx}
\usepackage{textcomp}
\usepackage{bm}
\usepackage{algorithm}
\usepackage{algorithmicx}
\usepackage{algpseudocode}
\usepackage{amssymb}
\usepackage{tikz}
\usepackage{booktabs}
\usepackage{multirow}
\usepackage{authblk}
\usepackage [misc] {ifsym}
\usepackage{cite}
\usepackage{xcolor}
\usepackage{changes}
\usepackage{subfig}
\usepackage{hyperref}

\usetikzlibrary{shapes.geometric, arrows}
\floatname{algorithm}{Algorithm}

\begin{document}
\title{A double-layer placement algorithm for integrated circuit-based modules on printed circuit board}
\author{Hangyuan Li,
Zhaoyang Yang,
Haotian Pang,
Ning Xu,Yu Chen
\thanks{H. Li and Z. Yang are with the School of Computer Science and Artificial Intelligence, Wuhan University of Technology, Wuhan, 430070, China.}
\thanks{H. Pang and N. Xu are with the School of Information Engineering, Wuhan University of Technology, Wuhan, 430070, China.}
\thanks{Y. Chen is with the School of Mathematics and Statistics, Wuhan University of Technology, Wuhan, 430070, China. Email: ychen@whut.edu.cn (Corresponding author)}
\thanks{This work is partially supported by the National Nature Science Foundation of China (No. 92373102), and partially supported by the Fundamental Research Funds for the Central University (No. 104972024KFYjc0055).}
}

\maketitle

\begin{abstract}
Considering that the physical design of printed circuit board (PCB) follows the principle of modularized design, this paper proposes an automatic placement algorithm for functional modules. We first model the placement problem as a mixed-variable optimization problem, and then, developed tailored algorithms of global placement and legalization for the top-layer centralized placement subproblem and the bottom-layer pin-oriented placement subproblem. Numerical comparison demonstrates that the proposed mixed-variable optimization scheme can get optimized total wirelength of placement. Meanwhile, experimental results on several industrial PCB cases show that the developed centralized strategies can well accommodate the requirement of top-layer placement, and the pin-oriented global placement based on bin clustering contributes to optimized placement results meeting the requirement of pin-oriented design.


\end{abstract}


\section{Introduction}
\label{sec:introduction}
As the carrier of components and circuits, printed circuit board (PCB) plays a critical role in the design and fabrication of consumer electronics. Due to the integration of diverse circuits with different characteristics, the task of PCB placement must follow diverse design rules and constraints of various PCB application scenarios. Consequently, component placement on PCB usually adopt a principle of modularized design, which is a challenging task that requires scenarios-oriented design of efficient algorithm.

The task of component placement is popularly implemented in two ways \cite{chang2009essential}. The approach based on a representative structure codes the positional relationships of the components into a discrete data structure such as the sequence-pair \cite{sequence-pairs}, the B*-trees \cite{B*-trees}, and the O-tree \cite{O-Tree}, and formulates it as a combinatorial optimization problem, which is typically addressed by metaheuristics \cite{metahr,memetic,sa}. The coordinate-based approach models the automatic placement as an optimization problem taking the component positions as decicion variables, which is usually solved by gradient-based analytical algortihms \cite{viswanathan2004fastplace,kahng2005aplace,chan2005multilevel,6269969}. To address the challenge that the gradient-based algorithm requires derivable objective function to be optimized, the subgradient methods are introduced~\cite{zhu2015nonsmooth,main_ref}. 

Since these function modules typically compose of an integrated circuit and some attached components, the two-layer placement task of function module consists of the top-layer centralized placement subproblem and the bottom-layer pin-oriented placement subproblem.
The approach based on a representative structure cannot accommodate it efficiently, and we prefer to implementing it by analytic methods.
An analytic placement algorithm for PCB consists of two stages. The first stage is to perform the global placement, which tries to get optimal positions of modules with partial overlapping. The second stage is legalization, which aims to eliminate overlap and achieve placement results complying with the design rules.
Moreover, PCB placement requires flexible rotation of modules to well accommodate the irregular placement region.
To address this challenge, we propose an automatic placement algorithm for IC-based modules (PA-ICM) on PCB  by addressing the following issues.
\begin{itemize}
    \item At the stage of global placement, a mixed-variable optimization model is established to regulate both rotation angles and positions of modules, which are simultaneously optimized by a hybrid algorithm based on the distribution evolutionary algorithm based on a population of probability model (DEA-PPM)~\cite{Xu2023} and the conjugate subgradient algorithm (CSA)~\cite{main_ref}.
    \item To meet the design requirement of PCB, we propose an efficient legalization algorithm that can generate legal placement fullfilling diverse space rules between different modules.
    \item  The performance of our algorithm is further improved with a cluster-based optimization strategy, by which a cluster-wise inialization, a batch opitmization strategy and the corresponding ligalization process contribute to placement results that meet the design rules of module placement.
\end{itemize}

The rest of this paper is organized as follows. Section \ref{sec:ReWork} reviews related publications on automatic placement. Section \ref{sec:pre} provides some preliminaries for this research. Then, the proposed PA-ICM is presented in Section \ref{sec:PA-ICM}. and numerical results are presented in Section \ref{sec:exp} to validate the efficiency of PA-ICM. Finally, Section \ref{sec:con} concludes this paper.

\section{Related Work}\label{sec:ReWork}

Analytic methods are popularly employed in floorplanning/placement scenarios of very-larege scale integrated circuit (VLSI). In order to overcome the difficulty of the standard quadratic programming method, Nam \emph{et al.} \cite{nam2006fast} proposed an innovative fast hierarchical quadratic programming layout algorithm (HQP). Chen and Zhu \cite{6238405} proposed a multilevel algorithm for placement of standard cells and VLSI, which employs a nonlinear programming technique and a best-choice clustering algorithm to take a global view of the whole netlist and placement information, and then uses an iterative local refinement technique during the declustering stage to further distribute the cells and reduce the wirelength. By modeling the placement task as an optimzation problem constrained by the Poisson equation, Lu \emph{et al.}  \cite{Lu2015} developed the analytic algorithm \emph{ePlace} for the cell placement of VLSI, and  Li \emph{et al.} \cite{li2022pef} proposed an efficient large-scale floorplanning algorithm for the task of large-scale floorplanning with fixed-outline. To address the challenge of developing a faster mixed-size placer without hardware acceleration and loss of solution quality, Peng and Zhu \cite{10266769} proposed a mixed-size placement algorithm based on a novel definition of potential energy  and a fast approximate computation scheme for partial derivatives of the potential energy for the Poisson’s equation.

 Since the anlaytic global placement cannot eliminate overlaps between modules, the legalization process is always introduced at the following stage.
 Peter \emph{et al.}~\cite{Peter2008} proposed a legalization algorithm for placement in VLSI, in which standard cells can be aligned to rows. Lin \emph{et al.}~\cite{Lin2016} use polygon to approximate the curve of the area of every soft module and solve an ILP for legalization, which is time consuming and is difficult to scale up for large-scale floorplanning problem. Moffitt \emph{et al.}~\cite{Moffitt2006} proposed the repair method \emph{Floorist} based on constraint graphs, where the insertion of constraints could result in dense constraint graphs that is expensive to visit all arcs. Li \emph{et al.} \cite{li2022pef} improved the graph-based legalization method based on \emph{Floorist}, where both insertion and deletion of constraints are performed to legalize the floorplan efficiently. Sun \emph{et al.} \cite{main_ref} performed the legalization by optimizing the commposite objective with increasing weight of overlap violation, which contributes to faster convergence and small wirelength of the final result.

 Machine learning is also implemented in the floorplanning/placement scenarios. Hu \emph{et al.}~\cite{hu2020graph2plan} presented the Graph2Plan, which utilizes deep learning to automatically generate VLSI floorplans. The method works by modeling a floorplan problem as a graph structure and using a graph neural network (GNN) to learn how to generate an optimized layout from graph data. Vassallo and Bajada~\cite{10546526} explores the use of Reinforcement Learning (RL) to automate learning and optimize circuit placement. They propose a new reinforcement learning framework that utilizes adaptive reward mechanisms to improve the performance of layout algorithms, especially in large-scale and complex VLSI designs. Lin \emph{et al.}~\cite{10.1145/3316781.3317803} proposed an accelerated algorithm named as  DREAMPlace, which approximates the objective function in the traditional layout optimization problem by using the neural network model, and making full use of parallel computing power to accelerate the layout process. Mirhoseini \emph{et al.}~\cite{mirhoseini2020chip} solved the chip layout problem using Deep Reinforcement Learning (DRL), where each step of the layout is seen as an action, and the strategy function associated with the high dimensional space and complex structure of chip layout is approximated by a deep neural network (DNN). The experimental results show that the deep reinforcement learning method is superior to the traditional heuristic layout method in the optimization of multiple design objectives, which provides a new idea for VLSI design.


Flexible rotation of modules is critical for improvement of design quality in floorplanning/placement. In some works based on simulated annealing implementation of placement~\cite{sheng2012simulated}, The rotation Angle of each component (e.g., 0°, 90°, 180°, 270°) is discrete and can be adjusted by selecting a different rotation angle at each iteration. Besides its expensive complexity, the efficiency of simulated annealing algorithm largely depends on its neighborhood solution generation strategy, which may ledas to its inefficient exploration in the solution space. Another component rotation strategy is based on particle swarm optimization\cite{sun2006floorplanning,kumar2020review}, its implementation is relatively simple, fewer parameters, easy to understand and implement. However, when the solution space is large or very complex, the particles may fall into local optimal solutions, resulting in the final result not being globally optimal. In some cases, particle populations may not have enough diversity in the search space because some particles are clustered in a small area, which can cause the search to become limited.
Huang \emph{et al.}~\cite{Huang2023} employed an analytical method to optimize the variable module rotation degree driven by wirelength, which allows the modeules to be placed in desired orietations.  Sun \emph{et al.}~\cite{main_ref} proposed to address the discerete solution space of orietation by DEA-PPM, which provides an efficient framework for coevolution of both discerete orietation and continuous coordinate.

For automatic placement of PCB components, Wang \emph{et al.}\cite{10652855} discussed how to realize automatic PCB layout under multiple constraints, and Li \emph{et al.}\cite{10652825} proposed a centralized placement method based on the sequence pair representation.
Various types of algorithms are widely used in PCB placement. ML-related algorithms play a significant role in PCB placement.  Based on a reinforcement learning -based agent for layout inference and
fine-tuning and a large language model -based agent for interactive optimization, Chen \emph{et al.}\cite{Chen2025} developed a novel agent-based framework that automatically generates PCB layouts meeting industrial constraints
through user interactions. To accelerate the placement process of PCB, Zhang \emph{et al.} \cite{zhang2025cypress} proposed a scalable GPU-accelerated PCB placement method inspired by VLSI. It incorporates
tailored cost functions, constraint handling, and optimized techniques adapted for PCB layouts. Taking the netlist of the already placed and unplaced circuits as input and abstract them into graphs, Chen \emph{et al.} \cite{10617495} proposed a subgraph matching based reference placement algorithm to achieve PCB placement reuse, thereby improving placement efficiency.

\section{Preliminaries}
\label{sec:pre}
\subsection{Problem Statement}
The position of component $v_i$ is represented by its central coordinate ($x_i$,$y_i$), and its orientation is denoted by $r_i$, where $r_i=j\pi/2$ ($j=0,1,2,3$, $i=1,2,\cdots,n$). Thus, a placement of module can be representd by a combination of vectors $\{\bm{x}$, $\bm{y}$, $\bm{r}\}$, where $\bm{x}=(x_1,x_2,\cdots,x_n)$, $\bm{y}=(y_1,y_2,\cdots,y_n)$, $\bm{r}=(r_1,r_2,\cdots,r_n)$.

The module placement problem requires separate optimization of the top and bottom layouts, which incorporate different objective functions and constraints. The top layer placement strives to achieve a uniform distribution as much as possible while ensuring non-overlapping placement and reduced wirelength, and the placement problem of the top layer is formulated as:
\begin{equation}
\begin{array}{cl}
        \min & W(\bm{x,y})\\
            s.t.&
        \begin{cases}
            D (\bm{x, y, r}) = 0, \\
            B (\bm{x, y, r}) = 0,     
        \end{cases}
\end{array}
\end{equation}
where $W(x,y)$ is the total wirelength, $D(x,y,r)$ is the sum of overlapping area, and $B(x,y,r)$ is the length beyond the external contour.
It is transformed to a unconstrained optimization model
\begin{equation}\label{pro:con}
  \footnotesize  min\,\,f(\bm{x,y,r})=\alpha_1 W(\bm{x,y})+\beta_1\sqrt{D(\bm{x,y,r})}+\gamma_1 B(\bm{x,y,r}).
\end{equation}
where $\alpha_1$, $\beta_1$, $\gamma_1$ are parameters to be confirmed. Here, the square root of $D(\bm{x, y,r})$ is adopted to ensure that all indexes to be minimized are of the same dimension. Ignoring the region constraint, the placement problem of the bottom layer is formulated as
\begin{equation}\label{pro:uncon}
    min\,\,f(\bm{x,y,r})=\alpha_2 W(\bm{x,y})+\beta_2\sqrt{D(\bm{x,y,r})}.
\end{equation}
Thee items in models \eqref{pro:con} and \eqref{pro:uncon} are defined as follows.
\paragraph{Total Wirelength ($W(\bm{x,y})$)} The total wirelength is taken as the sum of half-perimeter wirelength (HPWL)
\begin{equation}
    W(\bm{x,y})=\sum\limits_{e\in E}(\max\limits_{vi\in e}x_i
    -\min\limits_{vi\in e}x_i+\max\limits_{vi\in e}y_i
    -\min\limits_{vi\in e}y_i).
\end{equation}
\paragraph{Sum of Overlapping Area ($D(\bm{x,y,r})$)} The sum of overlapping area is computed by
\begin{equation}
    D(\bm{x,y,r})=\sum\limits_{i,j}O_{i,j}(\bm{x,r})\times
    \sum\limits_{i,j}O_{i,j}(\bm{y,r}),
\end{equation}
where $O_{i,j}(\bm{x,r})$ and $O_{i,j}(\bm{y,r})$
represent the overlapping lengths of components $i$ and $j$
in the X-axis and Y-axis directions, respectively.
Denoting $\Delta_x(i,j)=|x_i-x_j|$, we know
\begin{equation}
  \scriptstyle  O_{i,j}(\bm{x,r})=
    \begin{cases}
      \scriptstyle  \min(\hat{w_i},\hat{w_j}), & \mbox{if }
  \scriptstyle 0\leq\Delta_x(i,j)\leq\frac{|\hat{w_i}-\hat{w_j}|}{2},\\
    \scriptstyle \frac{\hat{w_i}-2\Delta_x(i,j)+\hat{w_j}}{2}
    , & \mbox{if }\scriptstyle \frac{|\hat{w_i}-\hat{w_j}|}{2}<
    \Delta_x(i,j)<\frac{\hat{w_i}+\hat{w_j}}{2},\\
    \scriptstyle 0, & \mbox{if }\scriptstyle \Delta_x(i,j)\geq\frac{\hat{w_i}+\hat{w_j}}{2}.
    \end{cases}
\end{equation}
where $\hat{w_i}$ is confirmed by
\begin{equation}
    \label{wi}
    \hat{w_i}=
    \begin{cases}
        w_i, & \mbox{if } r_i\in\{0,2\}, \\
        h_i, & \mbox{otherwise.}
    \end{cases}
\end{equation}
Denoting $\Delta_y(i,j)$=$|y_i-y_j|$, we have
\begin{equation}
  \scriptstyle  O_{i,j}(\bm{y,r})=
    \begin{cases}
       \scriptstyle \min(\hat{h_i},\hat{h_j}), &  \mbox{if } \scriptstyle 0\leq\Delta_y(i,j)\leq\frac{|\hat{h_i}-\hat{h_j}|}{2}, \\
       \scriptstyle \frac{\hat{h_i}-2\Delta_y(i,j)+\hat{h_j}}{2}, & \mbox{if }\scriptstyle \frac{|\hat{h_i}-\hat{h_j}|}{2}<\Delta_y(i,j)<        \frac{\hat{h_i}+\hat{h_j}}{2},\\
       \scriptstyle 0, & \mbox{if }\scriptstyle\Delta_y(i,j)\geq\frac{\hat{h_i}+\hat{h_j}}{2}.
    \end{cases}
\end{equation}
where $\hat{h_i}$ is confirmed by
\begin{equation}
    \label{hi}
    \hat{h_i} =
    \begin{cases}
        h_i, & if\quad r_i\in\{0,2\}, \\
        w_i, & otherwise.
    \end{cases}
\end{equation}
\paragraph{Sum of Width beyond the External Contour ($B(\bm{x,y,r})$)} For top layer placement problems,
the positions of components require to meet the following constraints:
\begin{equation*}
    \begin{cases}
        0\leq{x_i-\hat{w_i}/2,x_i+\hat{w_i}/2\leq{W^*}}, \\
        0\leq{y_i-\hat{h_i}/2,y_i+\hat{h_i}/2\leq{H^*}}.
    \end{cases}
\end{equation*}
where $W^*$ and $H^*$ are the width and the height of the contour. Let
\begin{equation*}
\begin{array}{ll}
  \scriptstyle   b_{1,i}(\bm{x,r})=\max(0,\hat{w_i}/2-x_i), & \scriptstyle
     b_{2,i}(\bm{x,r})=\max(0,x_i+\hat{w_i}/2-W^*), \\
  \scriptstyle   b_{1,i}(\bm{y,r})=\max(0,\hat{h_i}/2-y_i), & \scriptstyle  b_{2,i}(\bm{y,r})=\max(0,y_i+\hat{h_i}/2-H^*).
\end{array}
\end{equation*}
where $\hat{w_i}$ and $\hat{h_i}$ are confirmed by $\eqref{wi}$ and $\eqref{hi}$, respectively. Consequently, $B(\hat{x,y,r})$ can be confirmed by
\begin{equation}
    B(\bm{x,y,r})=\sum_{i=1}^n\left(\sum_{k=1}^2b_{k,i}(\bm{x,r})+\sum_{k=1}^2 b_{k,i}(\bm{y,r})\right).
\end{equation}

\subsection{The Conjugate Sub-gradient Algorithm}
The conjugate sub-gradient algorithm (CSA) is an efficient analytic algorithm for optimization of analytic non-smooth prolems, which was introduced to address the placement problem~\cite{zhu2015nonsmooth} as well as the floorplanning probllem~\cite{main_ref} of VLSI. The CSA iteration process, presented in Algorithm \ref{alg:CSA}, is introduced in this paper to achieve the optimal position of modules.

\begin{algorithm}[htb]
	\caption{$\bm{u}^*=CSA(f,\bm{u}_0,k_{max},s_0)$}\label{alg:CSA}
\begin{algorithmic}[1]
    \Require {Objective function $f(\bm{u})$, Initial solution $\bm{u}_0$, Maximum iterations $k_{max}$, Initial step control parameter $s_0$;}
		\Ensure{Optimal solution $\bm{u}^*$;}
		$\bm{g}_0 \in \partial f(\bm{u}_0)$, $\bm{d}_0=\bm{0}$, $k \leftarrow 1$\;
		\While{termination-condition 1 is not satisfied}		
        \State calculated subgradient $\bm{g}_k \in \partial f(\bm{u}_{k-1})$;
		\State	calculate Polak-Ribiere parameters $\eta _k=\frac{\bm{g}_k^T (\bm{g}_k-\bm{g}_{k-1})}{||\bm{g}_{k-1}||_2^2}$;
		\State	computed conjugate directions $\bm{d}_k=-\bm{g}_k+\eta _k \bm{d}_{k-1}$;
		\State	calculating step size $a_k=s_{k-1}/||\bm{d}_k||_2$;
		\State	renewal solution $\bm{u}_k=\bm{u}_{k-1}+a_k \bm{d}_k$
 and updata $\bm{u}^*$;
		\State	update step control parameters $s_k=qs_{k-1}$\;
        \EndWhile
\end{algorithmic}		
	\end{algorithm}



\subsection{The Distribution Evolutionary Algorithm Based on A Population of Probability Model}
Besides the optimal location of modules, an optimal placement also requires the best orientation combination of modules. Accordingly, it is optimized by the distribution evolutionary algorithm based on a population of probability model (DEA-PPM) presented by Algorithm \ref{alg:DEA-PPM}~\cite{Xu2023}.
\begin{algorithm}
    \caption{DEA-PPM}
    \label{alg:DEA-PPM}
    \begin{algorithmic}[1]
        \Require $isbottom$, $netlist$, $components\{\bm{x,y,r}\}$, $icpins$, $IC$.
        \Ensure updated $components\{x,y,r\}$ $bestSingle$.
        \State Initialize the population fitness array $\bm{f}$ and get the optimal fitness $f_{best}$;
        \State set current generation $t\gets0$, the last generation $t_{max}\gets20$;
        \While{$t<t_{max}$}
            \State $\bm{{Q^{\prime}}}(t)=OrthExpQ(\bm{Q(t-1), P(t-1)})$;
            \State $\bm{{P^{\prime}}}(t)=OrthExpQ(\bm{Q(t), P(t-1)})$;
            \State $(\bm{P}(t),\bm{X,Y},s)=UpdateXY(P'(t),\bm{X,Y},s)$;
            \State $\bm{Q}(t)=RefineQ(\bm{{P^{\prime}}}(t),\bm{P}(t),\bm{{Q^{\prime}}}(t))$;
            \State t=t+1;
        \EndWhile
    \end{algorithmic}
\end{algorithm}

\section{The proposed placement algorithm}
\label{sec:PA-ICM}
\begin{figure}[!htp]
    \centering
    \includegraphics[width=0.4\textwidth]{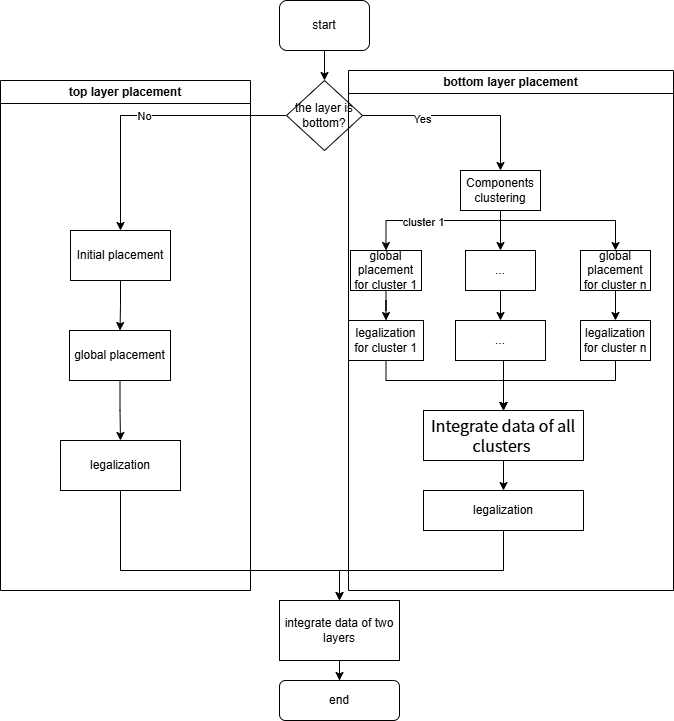}
    \caption{The Framework of the Proposed Method.}
    \label{fig:framework}
\end{figure}
To address the module placement of two-layer PCB, we propose a placement algorithm illustrated in Fig. \ref{fig:framework}. The purpose of the initial placement is to find the promising coordinates for all components. Then, the global placement is implemented to optimize the coordinates and orientations of the components with a delicate tradeoff between the total wirelength and the constraint violation. Finally, the legalization procedure aims to eliminate the overlap and get the final solution.

For placement of the bottom layer, the component clustering is performed during the initial placement phase. This step divides the placement problem into a number of sub-problems. The subsequent global placement and legalization of each cluster operate independently, without interfering with each other. After the local placement is completed, we must integrate the coordinates and orientations of the components in the various clusters, and the legalization procedure finally outputs the legal placement result.



\subsection{Placement Initialization}
Since the PCB design includes a variety of electric constraints that cannot be simulated at the stage of placement, the automatic placement of PCB components is significantly different from the placement scenarios of VLSI. Accordingly, tailored strategies are developed for the cases of PCB placement.

\subsubsection{Coordinate Initialization for the Top Layer}

\begin{algorithm}[!htp]
	\caption{TopInitialization}
	\label{alg:topini}
	\begin{algorithmic}[1]
        \Require $netlist$, $components\{\bm{x,y,r}\}$ of the top layer.
        \Ensure initialized $components$ of the top layer.
		\State compute total area of top layer components $cellsArea$ and approximate contour control
        parameter $k$;
        \State $totalArea \gets k\times cellsArea+icArea$;
        \State equal to the aspect ratio of IC, compute four vertices' coordinates of the external contour;
        \For{$i=1,\cdots,n$}
            \State $minDis\gets IntMax$, $md\gets components[i];$
            \For{$j=1,\cdots,n_i$}
                \State $net_j \gets nets[j];$
                \If{$net_j$ is a $closetopin$ net}
                    \State $p \gets$ the only pin in $\,net_j;$
                    \State $md\{x,y\} \gets p\{x,y\};$
                    \State $break;$
                \EndIf
                \State $ps\{p_1, p_2,\cdots, p_t\} \gets$ all pin points of $net_j$;
                \For{$k=1,\cdots,t$}
                    \State compute the distance of $p_k$ from the four edges of IC;
                    \State $dis \gets$ the minimum value of four values;
                    \If{$dis\leq minDis$}
                        \State $minDis \gets dis$, $md\{x,y\} \gets p_k\{x,y\}$;
                    \EndIf
                \EndFor
            \EndFor
                \State compute the distance of $md$ from the four edge of IC;
                \State select the direction with the smallest distance value to move $md$ outside the IC;
        \EndFor
	\end{algorithmic}
\end{algorithm}
Modules placed on the top layer, including an IC component and some accessary components, requires a centralized layout where accessary components are placed surounding the IC component. To meet this requirement, the coordinate initialization for components placed on the top layer is implemented as Algorithm \ref{alg:topini}, which consists of
three aspects: creating an artificial external contour, selecting the nearest associated pin point to connect for each component, and moving components to the valid positions. In order to control the density of placement and balance the number of components in different partitions, a moderately sized contour is built by lines 1-3. Then, we
can extract from the netlist data the connection relationship between components and IC-pins. After filtering out associated nets for each component, it is
convenient to compute the distance between each IC-pin and each edge of IC by lines 13-21. Eventually, the coordinate is consistent with the nearest IC-pin point. Meanwhile, it is noteworthy that some nets are denoted to be close to pin, which means that they have highest placement priority for the reason of electrical factors.  When the algorithm have met such a net, the component coordinate should be initialized directly by lines 8-12. Accordingly, each component are attached to a desired IC-pin point, and we can legalize the coordinate ensuring that components locate at the outside of IC by lines 22-23.



\subsubsection{Coordinate Initialization for the Bottom Layer}


According to the typical design of an IC component, we first cluster the pins to be connected to get nice placement, which is performed by
Algorithm \ref{alg:botini} based on the density-based spatial clustering of applications with noise (DBSCAN) algorithm~\cite{Ester1996}. However, due to the limitations of the algorithm itself, even the most accurate control parameters may result in a small number of pin points being classified as noise points. Our solution is to exclude components that have selected this type of
IC pins from the subsequent global placement and place them directly on that pin. However, for other components, they will be grouped with their selected IC-pins in the same cluster.

The DBSCAN-based initialization process can be roughly divided into three phases. Firstly, we randomly select attachment points for each component according to its connection relationship by lines 1-7. Considering that there is often a large free placement space around the noise points, the second step implemented by lines 8-14 is dedicated to reducing the overlap area of the global placement. The desired effect is that components belonging to the same net as the noise points can be attached to them as much as possible. The last stage is to meet the basic needs of industry. We will select to move the components with the close-to-pin identifier to its corresponding pin points by lines 15-21.

\begin{algorithm}[!htp]
    \caption{DoDBSCAN}
    \label{alg:botini}
    \begin{algorithmic}[1]
        \Require $netlist$, $components\{\bm{x, y}\}$ in the bottom layer, $icpins$.
        \Ensure initialized $components$ of the bottom layer, clustering results of components.
        \State Extract all the pin points $ps$ that the bottom layer component needs to be connected to by referring to the netlist.
        \State Confirm $epsilon$ and $minpts$, make the coordinates of $ps$ the objective to cluster.
        \For{$i=1,\cdots,n$}
            \State $md\gets components[i]$;
            \State Extract all the IC-pins $pts_i$ that belong to the same net as $md$;
            \State Randomly select a pin $p$ in $pts_i$ as the attachment point for $md$,
            $md\{x,y\}\gets p\{x,y\}$;
        \EndFor
        \State $np\{p_1,\cdots,p_k\} \gets$ all noise points in IC-pins;
        \For{$p_i \in np$}
            \If{$p_i$ is not selected by other components}
                \State $net \gets$ the net where $p_i$ is located;
                \State Randomly select a component in $net$ to attach to $p_i$;
            \EndIf
        \EndFor
        \State $closeNets \gets$ all closetopin nets in the bottom layer;
        \For{$net \in closeNets$}
            \State $p \gets$ the only pin in $net$;
            \If{$p$ is not selected}
                \State move the component of $net$ to $p$;
            \EndIf
        \EndFor
    \end{algorithmic}
\end{algorithm}

\subsection{Global Placement}

Algorithm \ref{alg5} presents the framework of global placement. It starts with initialization of
the orientation distribution and solution population $\bm{Q(0)}$ and $\bm{P(0)}$, then by checking the flag
$isbottom$, $components$' coordinates are initialized by lines 2-6. For the bottom layer placement, extra data structures like $netvec$ and $clusterIndex$ are also recorded. Since optimization algorithm has inevitable randomness, we attempt to set mulitple rounds optimization by lines 9-13 and choose the best solution as the output results.

\begin{algorithm}[!htp]
    \caption{Global Placement}
    \label{alg5}
    \begin{algorithmic}[1]
        \Require $isbottom$, $netlist$, $components\{x,y,r\}$, $icpins$, $IC$.
        \Ensure Refined $components\{x,y,r\}$ $bestSingle$.
        \State initialize $\bm{Q(0)}$ and generate $\bm{P(0)}$ by sampling $\bm{Q(0)}$;
        \If{$isbottom$==true}
            \State DoDBSCAN($netlist,components\{\bm{x,y,r}\}$);
        \Else
            \State TopInitialization($netlist, components\{\bm{x,y,r}\}, IC$);
        \EndIf
        \State let $\bm{(x^*,y^*,p^*)}$ = arg $\min{f(\bm{x,y,r})}$, $\{\bm{x,y,r}\}$ is obtained from $components$;
        \State set $\bm{q^*}$ as the distribution $\bm{q}$ corresponding to $\bm{p^*}$;
        \State $i\gets0$, $i_{max}=10$;
        \While{$i<i_{max}$}
            \State ResetXY$(components\{\bm{x,y}\})$;
            \State DEA-PPM();
        \EndWhile
    \end{algorithmic}
\end{algorithm}

To get the optimized result of global placement, the \emph{ResetXY} and \emph{DEA-PPM}, presented by Algoirthms \ref{alg6} and \ref{alg:DEA-PPM}, are iteratively implemented for $i_{max}$ times.
In addition to re-initializing the components' coordinates, \emph{ResetXY} also needs to specify parameters to the objective function of CSA. Since the amount of components in different clusters may vary greatly, clusters with a larger number of components often need to assign a larger overlap weight, so lines 4-9 provide an adaptive parameter allocation method.

The \emph{DEA-PPM} presented by Algorithm \ref{alg:DEA-PPM} simultaneously optimizes orientations and coordinates of components, where the coordinate update implemented by \emph{UpdateXY} (Algorithm \ref{alg:updatexy}) is consecutively run for $k$ clusters. Note that the cluster number $k$ is set as $1$ for the top layer and confirmed by the DBSACN for the bottom layer.

%
\begin{algorithm}[!htp]
    \caption{ResetXY}
    \label{alg6}
    \begin{algorithmic}[1]
        \Require $netlist,components\{\bm{x,y,r}\}$;
        \Ensure initialized $components\{\bm{x,y,r}\}$;
        \If{$isbottom$ == true}
            \State $alpha$$\gets$\{0,$\cdots$,0\}, $bta$$\gets$\{0,$\cdots$,0\};
            \State DoDBSCAN($netlist,components\{\bm{x,y,r}\}$);
            \For{$i=1,\cdots,k$}
                \State size $\gets$ the amount of components in cluster $i$;
                \State $\alpha\gets$ 1, $\beta\gets$ 1;
                \State $\beta\gets$ (size-10)$\div$10;
                \State $alpha[i]\gets\alpha$, $bta[i]\gets\beta$;
            \EndFor
        \Else
            \State TopInitialization($netlist, components\{\bm{x,y,r}\}, IC$);
            \State set $\alpha=1,\beta=20,\gamma=100$;
        \EndIf
        \State set step control parameter $css0$=100;
        \State reinitialize $\bm{Q(0)}$ and regenerate $\bm{P(0)}$ by sampling $\bm{Q(0)}$;
    \end{algorithmic}
\end{algorithm}

\begin{algorithm}[!htp]
    \caption{$(\bm{P}(t),\bm{X,Y},s)=UpdateXY(P'(t),\bm{X,Y},s)$}
    \label{alg:updatexy}
    \begin{algorithmic}[1]
        \Require $netvec,components\{\bm{x,y,r}\},icpins$;
        \Ensure Optimal individual $bestSingle$, optimal fintness $bestf$;
        \For{$i=1,\cdots,k$}
            \State $\alpha\gets alpha[i]$, $\beta\gets bta[i];$
            \State $cluster\gets components$ in cluster $i$;
            \State $net\gets netvec[i]$;
            \State $ps\gets icpins$ in cluster $i$;
            \State optimized ${cluster}=CSA(f,\bm{u}_0,k_{max},s_0)$;
            \State compute $HWPL, overlap$ based on current $cluster$ and $net$;
            \State $f=\alpha\times HPWL+\beta\times\sqrt{overlap}$;
            \If{$f<bestf_i$}
                \State $bestSingle_i\gets cluster$;
                \State $bestf_i \gets f$;
            \EndIf
        \EndFor
        \State $bestSingle \gets$ Integrate all local optimal $bestSingle_i$;
        \State $bestf \gets$ Add up all local optimal $bestf_i$;
    \end{algorithmic}
\end{algorithm}


\subsection{Legalization}
\label{legalization}

\begin{figure}[!htp]
  \centering
  \includegraphics[width=0.2\textwidth]{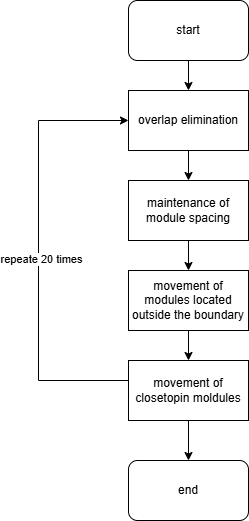}
  \centering
  \caption{The Flow Chart of the Legalization Process.}
  \label{legalize}
\end{figure}

The legalization process consisting of four steps is presented in Fig. \ref{legalize}. In the algorithm, eliminating the overlap function ensures that the placement complies with the basic design specification, maintaining spacing  is used to adjust the component spacing according to the prescribed spacing value in advance, the out-of-bounds component processing function is to re-locate the components moved out of the IC boundary by the previous two steps, and the close-to-pin components placement function is to force some components with special requirements to be placed near the pins.

\subsubsection{Legalization for the bottom layer}

\paragraph{Elimination of overlap}
\color{black}
The basic idea of eliminating overlap is to move the component up, down, left and right in four directions based on the component in the center position until there is no overlap.
Take the direction from the middle down as an example. The algorithm first determines an intermediate component $middle$ according to the component coordinates, and then continuously traverses the lower components from $middle$. For each traversed components, the algorithm will determine whether it overlaps with the upper component. If so, the algorithm will tentatively shift 1 unit downward until the current component does not overlap.

\paragraph{Maintenance of component spacing}
Eliminating overlap is to ensure that the algorithm meets the most basic design specifications. However, in various circuits, there are often larger spacing requirements between components. The algorithm adjusts the component coordinates according to the spacing file given in advance.

\paragraph{Out-of-bounds components processing}
In some placement results, there exist some cases in which several components are forced to move out of the IC boundary due to the influence of previous overlap elimination and the maintenance of the spacing function. The solution is to obtain the maximum blank space in the four corners of IC by computing the maximum rectangle algorithm in the bar chart, and then place the components within this region.

\paragraph{Movement of closetopin components}
Due to electrical and performance reasons, the netlist sometimes contains closetopin nets, which specify that a specified component must be placed around a specified pin point.
However, the legalization process could make some close-to-pin components deviating from its pin points.
Therefore, after the previous legalization steps, we will recheck the distances between the close-to-pin components and its pin points. If some distance exceeds our given threshold, the component will regenerate the coordinates around the specified pin point. The process of generating coordinates here is to randomly generate coordinate points in a two-dimensional normal distribution with a pin point as the center and a standard deviation of two.

\subsubsection{Legalization for the top layer}
\begin{figure}[htbp]
  \centering
  \includegraphics[width=0.3\textwidth]{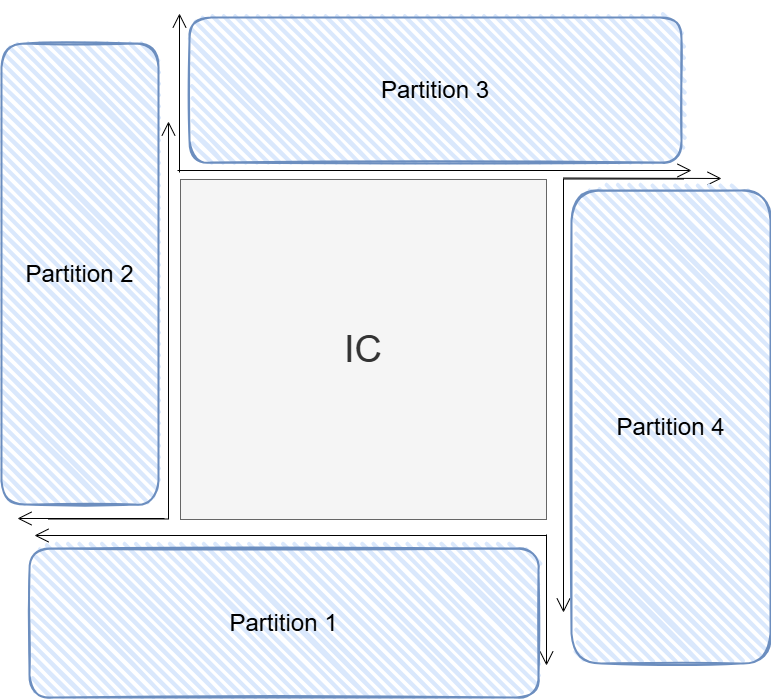}
  \centering
  \caption{Region Partitions and the Corresponding Overlapping Elimination Directions}
  \label{partition}
\end{figure}

The basic idea for component legalization at the top layer is consistent with that for the bottom layer. The difference is that the legalization for the top layer is accomplished based on partitions. Fig. \ref{partition} shows the regions in which the partitions are located, with the arrows in each partition indicating the direction in which the components within that partition eliminate overlap.

\section{Experimental Results and Analysis}
\label{sec:exp}
In order to verify the performance of our algorithm, we carefully selected four representative PCB placement examples and conducted experiments to obtain its performance indicators. Within these four cases, two are instances of two-layer PCB placement, one is a complete bottom-layer placement instance, and one is a complete top-layer placement.

All experiments are developed in C++ programming language program, and run in Microsoft Windows 11 on a laptop equipped with the AMD Ryzen 7 5800H @ 3.2GHz and 16GB system memory.

\subsection{Performance Comparison for the Floorplanning Benchmarks}
To highlight the superiority of our algorithm for optimization of wirelength, we compare the performance of rout algorithm with selected floorplanning algorithms via the selected floorplanning benchmarks.
Similar to our bottom layer placement scenario, they both refine the placement result within a fixed outline. Since the proposed algorithm is developed for PCB placement without soft modules, we only focuse on floorplanning algorithms developed for hard module floorplanning problems, including Parquet 4.5 \cite{parquet} and IARFP \cite{IARFP}.
The results of $30$ independent runs are included in Table \ref{res:fp}.
\begin{table}[!htp]
\caption{Numerical comparison for the fixed-outline floorplanning benchmarks with $\gamma$=15\%.}
\label{res:fp}
\centering
\resizebox{\columnwidth}{!}{
\begin{tabular}{|c|c|ccc|ccc|ccc|}
\hline
\multirow{2}{*}{circuit} & \multirow{2}{*}{aspect ratios} & \multicolumn{3}{c|}{Parquet 4.5}                                     & \multicolumn{3}{c|}{IARFP}                                           & \multicolumn{3}{c|}{Ours}                                           \\ \cline{3-11}
                         &                                & \multicolumn{1}{c|}{\#succ} & \multicolumn{1}{c|}{HPWL}    & time(s) & \multicolumn{1}{c|}{\#succ} & \multicolumn{1}{c|}{HPWL}    & time(s) & \multicolumn{1}{c|}{\#succ} & \multicolumn{1}{c|}{HPWL}   & time(s) \\ \hline
\multirow{3}{*}{ami33}   & 1.0                            & \multicolumn{1}{c|}{18\%}   & \multicolumn{1}{c|}{88689}   & 2.19    & \multicolumn{1}{c|}{90\%}   & \multicolumn{1}{c|}{90640}   & 2.72    & \multicolumn{1}{c|}{70\%}   & \multicolumn{1}{c|}{117746} & 0.58    \\ \cline{2-11} 
                         & 1.5                            & \multicolumn{1}{c|}{36\%}   & \multicolumn{1}{c|}{99106}   & 2.18  & \multicolumn{1}{c|}{98\%}  & \multicolumn{1}{c|}{93263}   & 4.18    & \multicolumn{1}{c|}{70\%}   & \multicolumn{1}{c|}{121308} & 0.53    \\ \cline{2-11}
                         & 2.0                            & \multicolumn{1}{c|}{10\%}   & \multicolumn{1}{c|}{99092}   & 2.17    & \multicolumn{1}{c|}{70\%}   & \multicolumn{1}{c|}{93590}   & 4.19    & \multicolumn{1}{c|}{45\%}   & \multicolumn{1}{c|}{127489} & 0.76    \\ \hline
\multirow{3}{*}{ami49}   & 1.0                            & \multicolumn{1}{c|}{50\%}   & \multicolumn{1}{c|}{1205430} & 5.39    & \multicolumn{1}{c|}{86\%}   & \multicolumn{1}{c|}{1065665} & 5.31    & \multicolumn{1}{c|}{100\%}  & \multicolumn{1}{c|}{870427} & 0.31    \\ \cline{2-11}
                         & 1.5                            & \multicolumn{1}{c|}{54\%}   & \multicolumn{1}{c|}{1227390} & 5.37    & \multicolumn{1}{c|}{90\%}   & \multicolumn{1}{c|}{1050709} & 5.32    & \multicolumn{1}{c|}{100\%}  & \multicolumn{1}{c|}{872344} & 0.41    \\ \cline{2-11}
                         & 2.0                            & \multicolumn{1}{c|}{50\%}   & \multicolumn{1}{c|}{1264920} & 5.42    & \multicolumn{1}{c|}{84\%}   & \multicolumn{1}{c|}{1109924} & 5.36   & \multicolumn{1}{c|}{100\%}  & \multicolumn{1}{c|}{892254} & 0.65    \\ \hline
\multirow{3}{*}{n100}    & 1.0                            & \multicolumn{1}{c|}{64\%}   & \multicolumn{1}{c|}{343496}  & 20.56   & \multicolumn{1}{c|}{100\%}  & \multicolumn{1}{c|}{312400}  & 7.07    &  \multicolumn{1}{c|}{100\%}  & \multicolumn{1}{c|}{292339} & 0.83    \\ \cline{2-11}
                         & 1.5                            & \multicolumn{1}{c|}{42\%}   & \multicolumn{1}{c|}{337230}  & 20.59   & \multicolumn{1}{c|}{100\%}  & \multicolumn{1}{c|}{313305}  & 7.34    & \multicolumn{1}{c|}{100\%}  & \multicolumn{1}{c|}{300342} & 0.97    \\ \cline{2-11}
                         & 2.0                            & \multicolumn{1}{c|}{40\%}   & \multicolumn{1}{c|}{346051}  & 20.38   & \multicolumn{1}{c|}{100\%}  & \multicolumn{1}{c|}{310485}  & 7.48    & \multicolumn{1}{c|}{100\%}  & \multicolumn{1}{c|}{308472} & 1.06    \\ \hline
\multirow{3}{*}{n200}    & 1.0                            & \multicolumn{1}{c|}{60\%}   & \multicolumn{1}{c|}{653990}  & 88.97   & \multicolumn{1}{c|}{100\%}  & \multicolumn{1}{c|}{561799}  & 20.75  & \multicolumn{1}{c|}{100\%}  & \multicolumn{1}{c|}{521598} & 2.31    \\ \cline{2-11}
                         & 1.5                            & \multicolumn{1}{c|}{30\%}   & \multicolumn{1}{c|}{648938}  & 91.46   & \multicolumn{1}{c|}{100\%}  & \multicolumn{1}{c|}{571687}  & 21.11   & \multicolumn{1}{c|}{100\%}  & \multicolumn{1}{c|}{529829} & 2.52    \\ \cline{2-11}
                         & 2.0                            & \multicolumn{1}{c|}{30\%}   & \multicolumn{1}{c|}{678282}  & 88.59   & \multicolumn{1}{c|}{100\%}  & \multicolumn{1}{c|}{559333}  & 22.37   & \multicolumn{1}{c|}{100\%}  & \multicolumn{1}{c|}{540140} & 2.64    \\ \hline
\multirow{3}{*}{n300}    & 1.0                            & \multicolumn{1}{c|}{80\%}   & \multicolumn{1}{c|}{798571}  & 189.97  & \multicolumn{1}{c|}{100\%}  & \multicolumn{1}{c|}{667628}  & 34.29  & \multicolumn{1}{c|}{100\%}  & \multicolumn{1}{c|}{587840} & 3.96    \\ \cline{2-11}
                         & 1.5                            & \multicolumn{1}{c|}{10\%}   & \multicolumn{1}{c|}{847515}  & 194.27  & \multicolumn{1}{c|}{100\%}  & \multicolumn{1}{c|}{658637}  & 36.63   & \multicolumn{1}{c|}{100\%}  & \multicolumn{1}{c|}{606776} & 3.94    \\ \cline{2-11}
                         & 2.0                            & \multicolumn{1}{c|}{10\%}   & \multicolumn{1}{c|}{915228}  & 191.06  & \multicolumn{1}{c|}{100\%}  & \multicolumn{1}{c|}{658410}  & 36.50   & \multicolumn{1}{c|}{`100\%} & \multicolumn{1}{c|}{627223} & 4.26    \\ \hline
\end{tabular}
}
\end{table}

Numerical results demonstrate that the proposed algorithm runs generally faster than Parquet 4.5 and IARFP. The combinatorial explosion results in fast increase of runtime and significant performance degeneration of Parquet 4.5 and IARFP, while the proposed algorithm employing the mixed-variable model performs reasonably well on the circuits $ami49$, $n100$, $n200$ and $n300$. However, for the small-scale circuit $ami33$, our algorithm cannot outperform than Parquet 4.5 and IARFP, because the comibinatorial optimization mechanisms can better accommodate the compact placement problem for small-scale floorplanning problems.

\subsection{Performance Validation for Industrial PCB Cases}
In order to validate the performance of our placement algorithm, four industrial cases presented in Table \ref{exp-info} are investigated. The first two cases are single layer placement problems, where Case 1 requires the bottom layer placement, and Case 2 is to place components on the top layer. Cases 3 and 4 include components on both layers, and they require automatic placement for both the top and the bottom layers.
\begin{table}[!htp]
\centering
\caption{The Circuit Parameters of Investigated PCB Cases.}
\label{exp-info}
\resizebox{\columnwidth}{!}{
\begin{tabular}{|l|l|l|l|}
\hline
Circuit   & Blocks(top layer + bottom layer) & Terminals(IC-Pins) & Total Nets \\ \hline
Case1    & 0+148                            & 1932               & 518        \\ \hline
Case2       & 80+0                             & 662                & 214        \\ \hline
Case3 & 19+61                            & 642                & 213        \\ \hline
Case4  & 39+104                           & 441                & 392        \\
\hline
\end{tabular}
}
\end{table}

\subsubsection{Validation of the Legalization Procedeure}
As presented before, we propose a flexible legalization method for the placement of PCB. According to the different placement principles of two layers, the proposed legalization procedure incorporates two different legalization strategies. Taking Case 3 as a instance of two-layer placement, we demonstrate how the legalization procedure transforms the result of global placement to a legalized placement.
\begin{figure}[!htp]
\label{leg}
\centering
  \subfloat[Pre-legalization placement]{\includegraphics[width=0.24\textwidth]{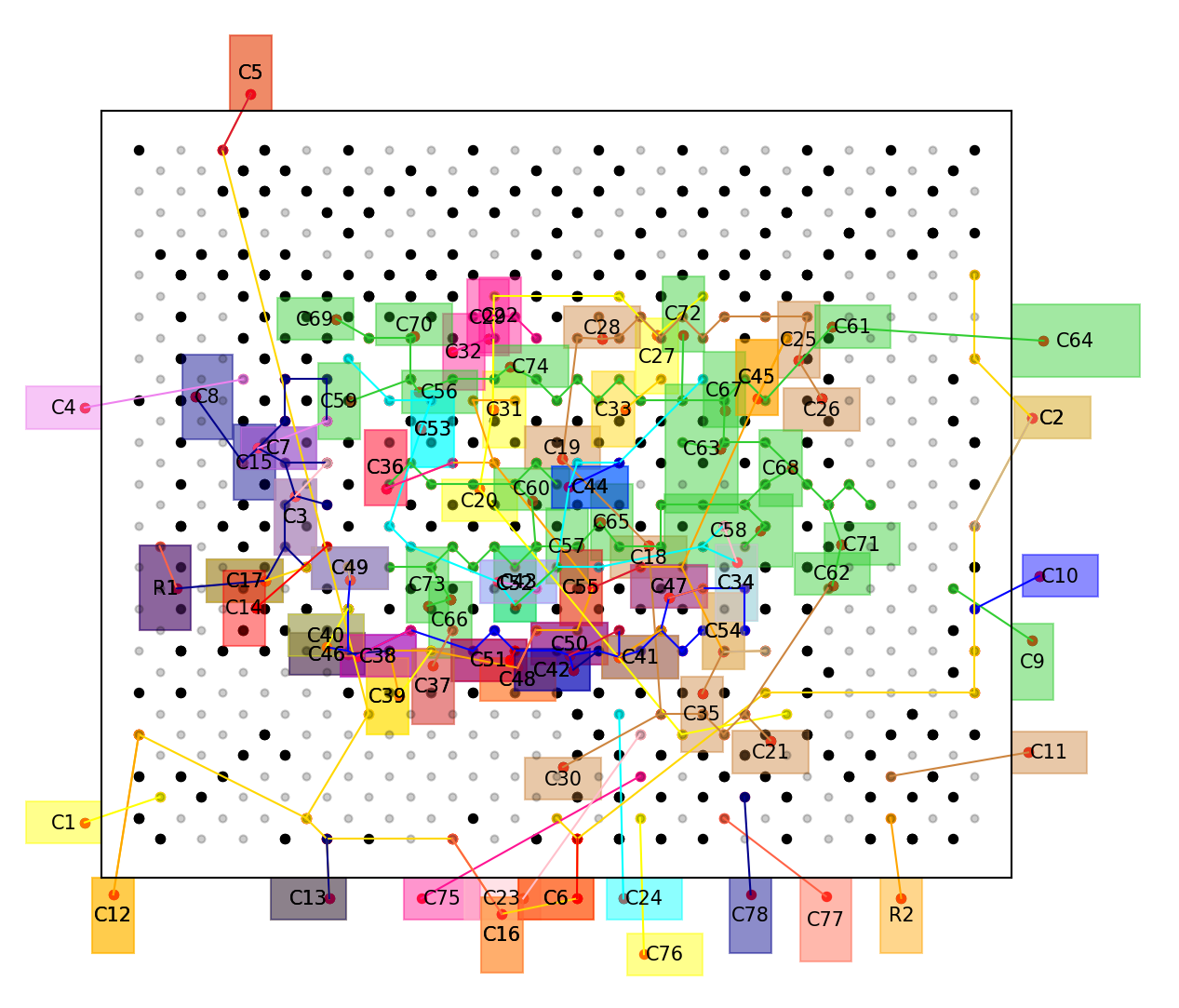}
      \label{leg-before-fig}}
\subfloat[Post-legalization placement]{\includegraphics[width=0.24\textwidth]{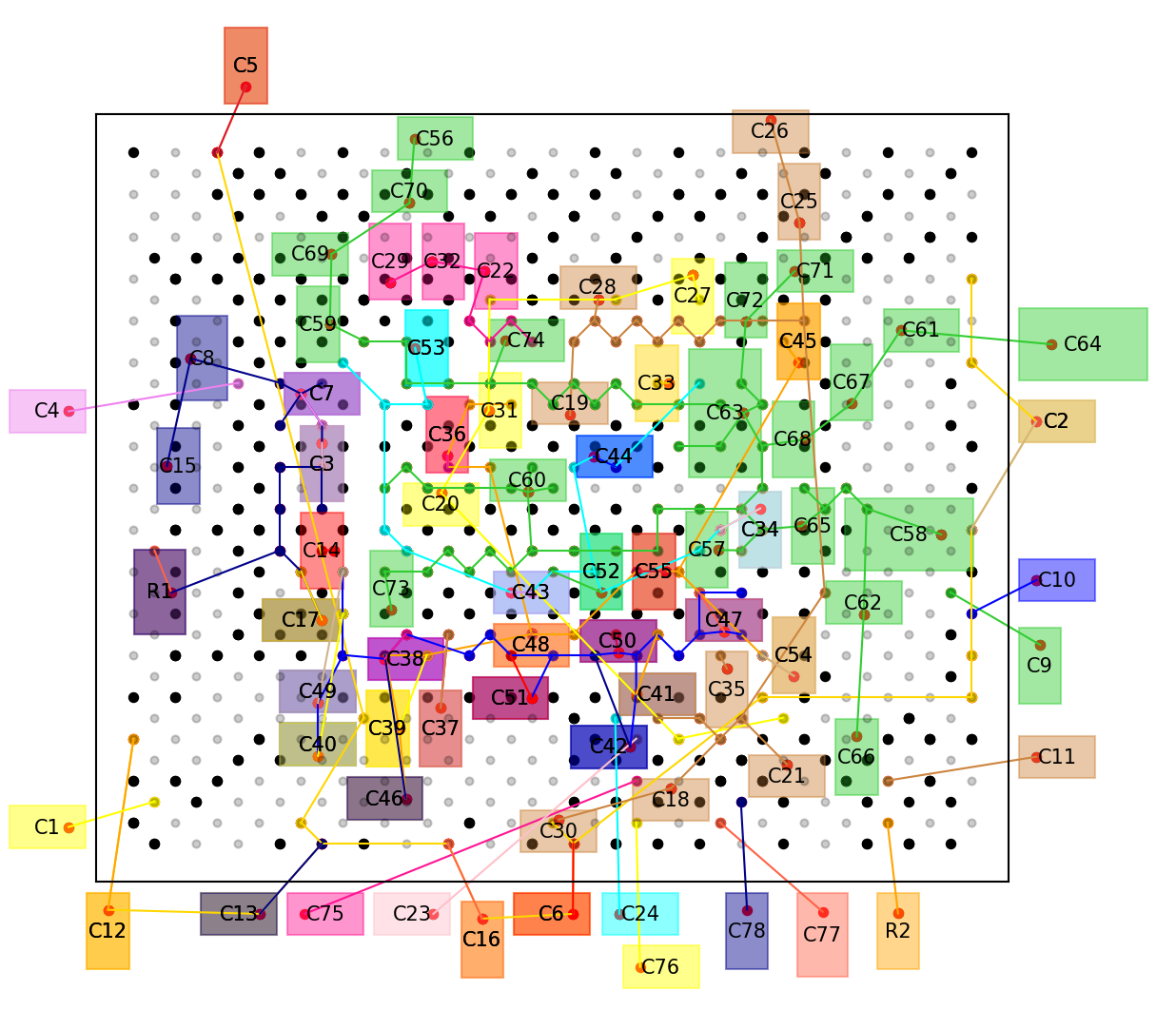}\label{leg-after-fig}}
\caption{An comparative illustration for pre- and post-legalization placement results.}\label{fig:leg}
\end{figure}

The comparative illustration for performance of legalization is included in Fig. \ref{fig:leg}. Taking the placement results illustrated by Fig. \ref{leg-before-fig} as the input, the legalization get the final placement presented by \ref{leg-after-fig}. Besides the elimination of overlap, the result of legalization generally meet the different principle of the two-layer PCB placement.
\begin{itemize}
  \item For centrally located components, it maintains a small offset distance and keeps a relative position relationship with surrounding components.
  \item The close-to-pin components (e.g. R1, C14 and C50) keeps  adjacent to the assigned pins.
  \item For the bottom-layer placement, all components are placed in the feasible region restricted by the outline of IC.
\end{itemize}

\subsubsection{Placement results of PCB}
\begin{figure}[!htp]
\centering
\subfloat[Case 1]{\includegraphics[width=0.48\columnwidth]{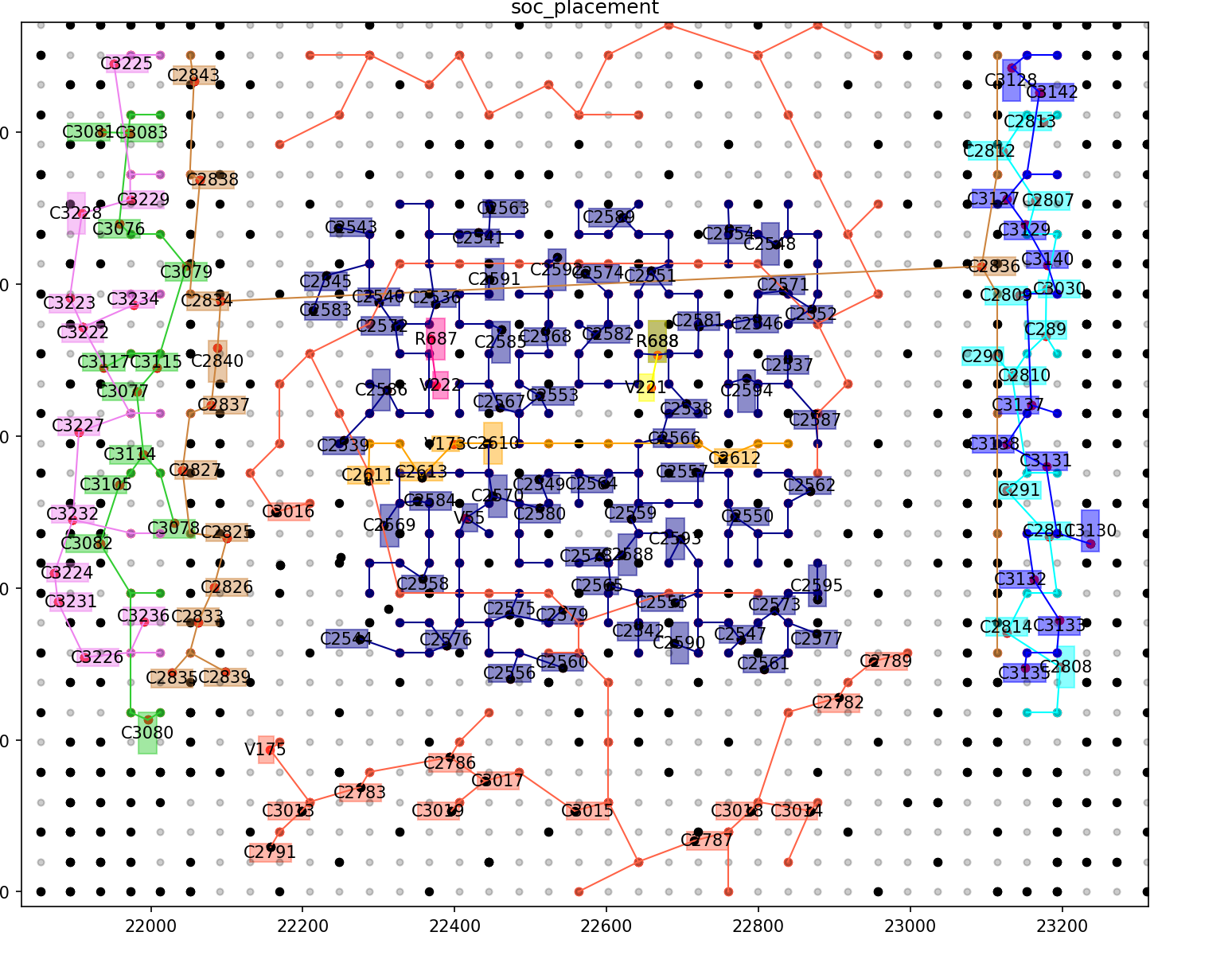}
      \label{A10_SI-fig}}
\subfloat[Case 2]{\includegraphics[width=0.48\columnwidth]{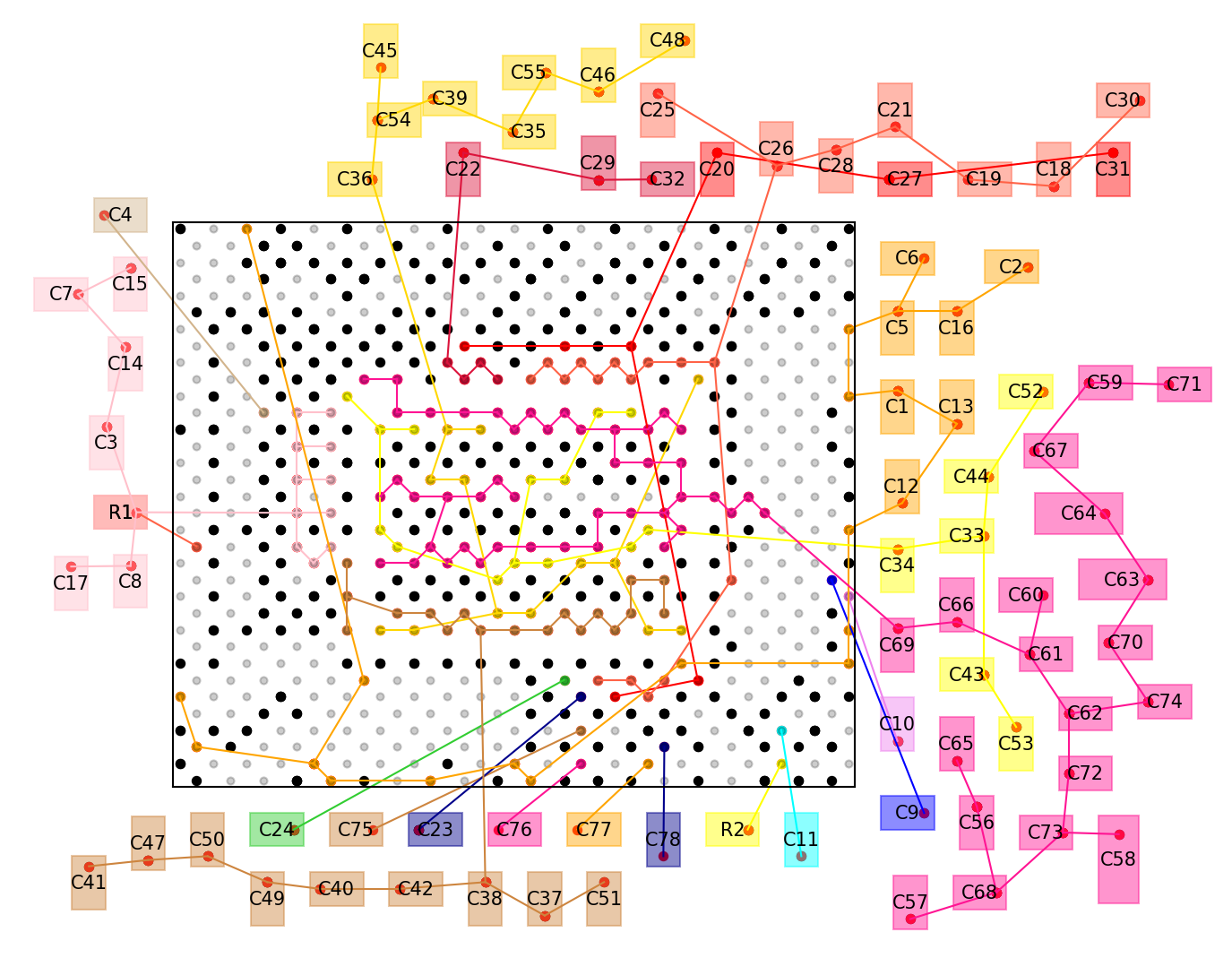}
      \label{Case2-fig}}\\
\subfloat[Case 3]{\includegraphics[width=0.48\columnwidth]{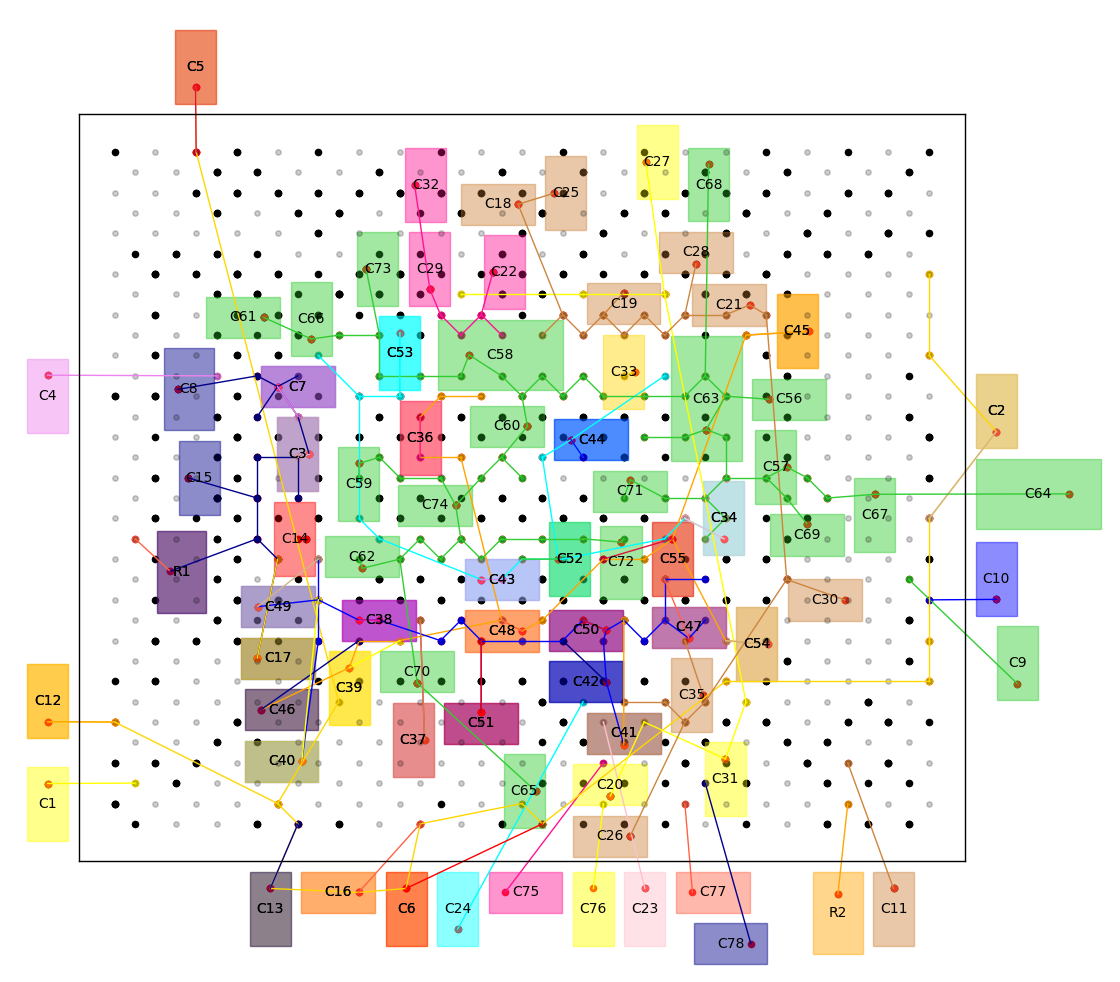}
      \label{case3-fig}}
\subfloat[Case 4]{\includegraphics[width=0.48\columnwidth]{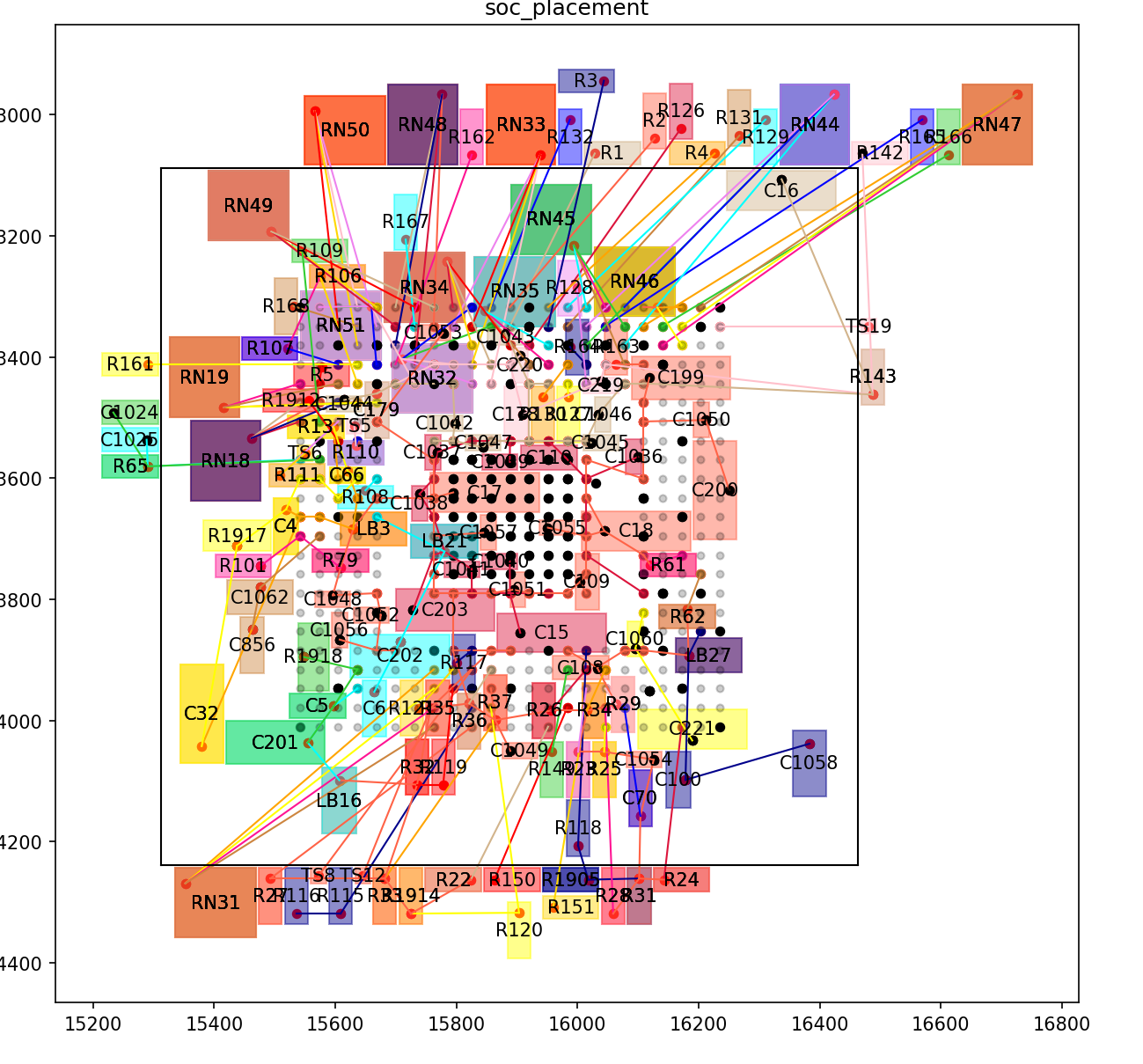}
      \label{Case4-fig}}
  \caption{The final placement results of the proposed algorithm.}
  \label{fig:pcb}
\end{figure}

The placement proposed in this paper are implemented to address the problems included in Table. \ref{exp-info}, and the obtained results are illustrated in Fig. \ref{fig:pcb}, which demonstrates that our proposed algorithm can well address both the top-layer and the bottom-layer placement.
For ten independent runs, the statistic results of HPWL and runtime are included in Table \ref{tab:pcb}, which shows that our algorithm can get the automatic placement of components with a couple of minutes. The small standard deviation values of HPWL and runtime also validate the performance stability of the proposed algorithm.

\begin{table}[!htp]
\centering
\caption{Statistical results of HPWL and runtime for the PCB cases.}
\label{tab:pcb}
{
\begin{tabular}{|c|c|c|c|c|c|}
  \hline
  \multirow{2}{*}{Circuit} & \multirow{2}{*}{Layer} & \multicolumn{2}{c|}{HPWL(mil)} & \multicolumn{2}{c|}{CPU(s)} \\
  \cline{3-6}
   & & mean & std & mean & std\\
   \hline
   Case1 & bottom & 9267.355 & 34.689 & 8.985 & 0.483\\
   \hline
   Case2 & top & 4667.297 & 56.118 & 3.1995 & 0.286 \\
   \hline
   \multirow{2}{*}{Case3} & top & 2414.195 & 112.226 & \multirow{2}{*}{4.165} & \multirow{2}{*}{0.211}\\
   \cline{2-4}
    & bottom & 2663.65 & 68.98 &  &\\ \hline
    \multirow{2}{*}{Case4} & top & 20570.730 & 241.731 & \multirow{2}{*}{32.702} & \multirow{2}{*}{1.816}\\
   \cline{2-4}
    & bottom & 7984.445 & 904.538 &  &\\
   \hline

\end{tabular}
}
\end{table}

\paragraph{The bottom-layer placement} The bottom layer placement is required for Case1, Case 3 and Case 4. 
Although it performs typically well for these cases, the standard deviation value of HPWL for Case4 is not extremely small, which shows that the stability could degrade when addressing the case that the bottom IC-pin points density is very concentrated. One reason is that the density-based clustering algorithm tends to cluster IC-pin points in a concentrated distribution into one cluster, which increases the challenge of global placement. In addition, the practice of splitting multiple netlists by clusters may affect the overall connectivity between the pin points of some nets.

\paragraph{The top-layer placement} For the top-layer placement of  Case2, Case3 and Case4, our algorithm can also arrange the components properly. Despite of the different characteristics of them, the proposed algorithm well addresses the unbalanced distribution of components (Case 2 and Case 4) and generates legal placement with various setting of component space. Considering that the top-layer components are preferred to being connected via surface routing, it achieves nice placement results because there are only a few number of crossed jumper wires.

\section{Conclusion}
\label{sec:con}
In this paper, we formulate the 2-layer module placement problem as a mixed-variable optimization problem, and develop tailored strategies for global optimization and legalization of both the top-layer placement and the bottom-layer placement. Numerical comparison demonstrates that the proposed mixed-variable optimization scheme can outperform the metaheuristics algorithm based on representative structure codes, and the placement results on industrial PCB cases show that the proposed placement algorithm for IC-based modules can get satisfactory results for real PCB design scenarios. Our future study will focus on improve its performance on large-scale cases, and incorporate the module placement method to implement automatic design of a PCB consisting of several function modules.

\bibliographystyle{IEEEtran}
\bibliography{IEEEabrv,references}

\end{document}